\documentclass
[superscriptaddress,secnumarabic,amssymb,amsmath,nobibnotes,aps,prd,showkeys,showpacs,nofootinbib,onecolumn]{revtex4}%
\usepackage{graphicx}
\usepackage{subfigure}
\usepackage{epstopdf}
\usepackage{epsf}
\usepackage{bm}
\usepackage{amsmath}
\usepackage{amsfonts}
\usepackage{amssymb}%
\providecommand{\U}[1]{\protect\rule{.1in}{.1in}}
\newcommand{\be}{\begin{equation}}
\newcommand{\ee}{\end{equation}}

\newcommand{\mincir}{\raise
-3.truept\hbox{\rlap{\hbox{$\sim$}}\raise4.truept\hbox{$<$}\ }}
\newcommand{\magcir}{\raise
-3.truept\hbox{\rlap{\hbox{$\sim$}}\raise4.truept\hbox{$>$}\ }}

\begin{document}
\title{Einstein-aether theory in Weyl integrable geometry}
\author{Andronikos Paliathanasis}
\email{anpaliat@phys.uoa.gr}
\affiliation{Institute of Systems Science, Durban University of Technology, Durban 4000,
South Africa}
\affiliation{Instituto de Ciencias F\'{\i}sicas y Matem\'{a}ticas, Universidad Austral de
Chile, Valdivia 5090000, Chile}
\author{Genly Leon}
\email{genly.leon@ucn.cl}
\affiliation{Departamento de Matem\'{a}ticas, Universidad Cat\'{o}lica del Norte, Avda.
Angamos 0610, Casilla 1280 Antofagasta, Chile.}
\author{John D. Barrow}
\email{J.D.Barrow@damtp.cam.ac.uk}
\affiliation{DAMTP, Centre for Mathematical Sciences, University of Cambridge, Wilberforce
Rd., Cambridge CB3 0WA, UK}

\begin{abstract}
We study the Einstein-aether theory in Weyl integrable geometry. The scalar
field which defines the Weyl affine connection  is introduced in the
gravitational field equation. We end up with an Einstein-aether scalar field
model where the interaction between the scalar field and the aether field has
a geometric origin. The scalar field plays a significant role in the evolution
of the gravitational field equations. We focus our study on the case of
homogeneous and isotropic background spacetimes and study their dynamical
evolution for various cosmological models.

\end{abstract}
\date{\today}
\maketitle

\section{Introduction}

\label{sec1}

Alternative theories of gravity \cite{clifton1}, where the Lorentz symmetry is
violated, have drawn the attention of gravitation physicists in the last
decades. Ho\v{r}ava-Lifshitz gravity and Einstein-aether gravity are two
theories which have been widely studied because they provide Lorentz violation.

Ho\v{r}ava-Lifshitz gravity is a power-counting renormalization theory with
consistent ultra-violet behavior exhibiting an anisotropic Lifshitz scaling
between time and space at the ultra-violet limit, while general relativity is
provided as a limit \cite{hor3}. There are various physical applications of
Ho\v{r}ava-Lifshitz gravity, some results on compact stars, black holes,
universal horizons, non-relativistic gravity duality and other subjects are
discussed in the review \cite{Wang:2017brl}. Recently, it has been found that
Ho\v{r}ava-Lifshitz gravity is in agreement with the observations of the
gravitational-wave event GW170817 \cite{gw01}\textbf{.} Cosmological
applications of Ho\v{r}ava-Lifshitz theory are discussed in
\cite{Cai:2009qs,Christodoulakis:2011np,Saridakis:2009bv,Kiritsis:2009sh,Lu:2009em}%
, while cosmological constraints on Ho\v{r}ava-Lifshitz\ theory using some of
the recent cosmological data can be found in \cite{Nilsson:2018knn}.

On the other hand, in Einstein-aether theory, the Lorentz symmetry is violated
by the introduction of a unitary time-like vector field, known as the
`\ae ther', in the Einstein-Hilbert action
\cite{Jacobson:2000xp,Carruthers:2010ii,Zlosnik:2006zu,Jacobson,Carroll:2004ai}%
. In particular, the terms quadratic in the covariant derivatives of the
kinetic part of the aether lagrangian modify the gravitational action of
general relativity. The Einstein-aether action is the most general
second-order theory which is defined by the spacetime metric $g_{ab}$ and the
aether field $u^{a}$ involves no more than two derivatives
\cite{Jacobson,Carroll:2004ai}. A main property of the Einstein-aether theory
is that it can be seen as the classical limit of Ho\v{r}ava-Lifshitz gravity.
This means that exact solutions of Einstein-aether theory are solutions of
Ho\v{r}ava-Lifshitz theory; however, the inverse is not true. While the two
theories are equivalent in terms of exact solutions, in general, the
equivalence is not true when the full field equations are considered
\cite{sot01}.

Black hole solutions in Einstein-aether gravity are discussed in \cite{bh01},
where it was found that the exterior solution is close to the Schwarzschild
solution of general relativity. On the other hand, the interior solution
differs from that of general relativity. In \cite{bh03}, the authors studied
spherically symmetric spacetimes with fluid source which describe neutron
stars. A detailed study of spherically symmetric spacetimes with perfect fluid
models in Einstein-\ae ther theory was performed in ref. \cite{col11}. The
authors studied the local stability of the equilibrium points for the
gravitational field equations for various perfect fluid models. The results of
\cite{col11} describe inhomogeneous cosmological models and astrophysical
objects, Kantowski-Sachs Einstein-\ae ther perfect fluid models were studied
in \cite{latta1}. In the presence of a scalar field, static spherically
symmetric solutions were determined in \cite{col112,col113}. Moreover, exact
solutions of homogeneous and anisotropic spacetimes were found recently in
\cite{roum1,roum2}. In addition, exact solutions in the presence of a modified
Chaplygin gas or in the presence of a Maxwell field can be found in
\cite{ch1,ch2} while the Einstein-aether theory has been studied as
dark-energy candidate to explain the late-time acceleration phase of the
universe in \cite{dea1,dea2}.

A Lorentz violating inflationary model has been proposed by Kanno and Soda in
\cite{Kanno:2006ty}. More precisely, it has been proposed a nonminimally
coupling of a scalar field with the aether field, where the Einstein-aether
coefficient becomes a function of the scalar field. In this model, the
inflationary stage is divided into two parts; the Lorentz-violating stage and
the standard slow-roll stage. In the first stage, the universe expands as an
exact de Sitter spacetime, although the inflaton field is rolling down the
potential. Another Einstein-aether scalar-field inflaton model coupled
bilinearly to the expansion of the aether was studied proposed by Donnelly and
Jacobson in \cite{DJ}. In Friedmann--Lema\^{\i}tre--Robertson--Walker (FLRW)
spacetime the inflaton-aether expansion coupling leads to a driving force on
the inflaton that is proportional to the Hubble parameter. This force affects
the slow-roll dynamics, but still allows a graceful exit of inflation. In
\cite{Barrow:2012qy}, several families of inflationary exact solutions were
derived for the model of Donnelly and Jacobson, while the effects of the
Lorentz violation during the slow-roll period were studied in \cite{bar2}.
Recently, homogeneous and isotropic exact solutions for an Einstein-aether
scalar field model were determined in \cite{ae1,ae2,ae3,ae4}, while studies of
homogeneous and anisotropic models in the Einstein-aether scalar field model
can be found in \cite{ans1,ans2,ans3}.

In this paper, we study the extension of Einstein-aether theory in Weyl
geometry, specifically in Weyl integrable theory \cite{cur0}. Weyl geometry is
a torsion-free manifold equipped with a connection which preserves the
conformal structure. In the case, where the conformal structure is analogous
to a scalar field, we have the Weyl Integrable theory where the connection
structure of the geometry differs from the Levi-Civita connection by a scalar
field of the conformal metric, which defines the Levi-Civita connection.
Hence, a scalar field can be introduced in the gravitational action integral
by geometric quantities of the underlying manifold. For various applications
of Weyl integrable theory in gravitational physics, we refer the reader to
\cite{va5,va6,va7,va2,va8,va9} and references therein. As we shall see from
the following analysis, in our approach we are able to introduce a geometric
scalar field coupled to the aether field, by considering the Einstein-\ae ther
action in Weyl geometry. The scalar field is introduced by the symmetric parts
of the covariant derivatives for the aether field. We consider the simplest
scenario of a homogeneous and isotropic spacetime, where we observe that the
scalar field is introduced by the expansion rate of the aether field. The
presence of an additional matter source is also discussed. The plan of this
paper is as follows.

In Section \ref{sec2}, we briefly discuss the main properties of Weyl geometry
and present the Einstein-Hilbert action in Weyl integrable theory. In Section
\ref{sec2a}, we present the model of our study, which is Einstein-aether
theory in Weyl geometry.\ We produce the gravitational field equations and we
focus on the case of a FLRW background space. We find that the field equations
can be solved explicitly, where the scale factor can describe an accelerated
universe for a specific range for the values of the free parameters of the
model. In Sections \ref{sec4} and \ref{sec5} we consider cosmological models
with a dust fluid source, or with a non-zero scalar field potential
respectively. Finally, in \ref{sec6} we discuss our results and we draw our conclusions.

\section{Weyl integrable gravity}

\label{sec2}

Consider a four-dimensional manifold $M$ described by the metric tensor,
$g_{\mu\nu}$, and the covariant derivative, $\tilde{\nabla}_{\mu}$, defined by
the affine connection $\tilde{\Gamma}_{\mu\nu}^{\kappa}$ with the property,
\begin{equation}
\tilde{\nabla}_{\kappa}g_{\mu\nu}=\omega_{\kappa}g_{\mu\nu}, \label{ww.01}%
\end{equation}
where $\omega_{\mu}~$is a gauge field which defines the geometry.

By definition (\ref{ww.01}), it follows that the affine connection
$\tilde{\Gamma}_{\mu\nu}^{\kappa}$ is related to the Christoffel
symbols~$\Gamma_{\mu\nu}^{\kappa}\left(  g\right)  $ of the metric tensor
$g_{\mu\nu}$ by the relation%
\begin{equation}
\tilde{\Gamma}_{\mu\nu}^{\kappa}=\Gamma_{\mu\nu}^{\kappa}-\omega_{(\mu}%
\delta_{\nu)}^{\kappa}+\frac{1}{2}\omega^{\kappa}g_{\mu\nu}, \label{ww.02}%
\end{equation}
from which we observe the importance of the gauge vector, $\omega_{\mu}$. When
$\omega_{\mu}$ is a gradient vector field, that is, when there exists a scalar
field, $\phi$, such that $\omega_{\mu}=\phi_{,\mu}$, then the Weyl geometry
reduces to the so-called Weyl integrable geometry.

In Weyl integrable geometry, the Ricci tensor $\tilde{R}_{\mu\nu}$ is related
to the Riemannian Ricci tensor $R_{\mu\nu}$ by \cite{salim96},%
\begin{equation}
\tilde{R}_{\mu\nu}=R_{\mu\nu}-\tilde{\nabla}_{\nu}\left(  \tilde{\nabla}_{\mu
}\phi\right)  -\frac{1}{2}\left(  \tilde{\nabla}_{\mu}\phi\right)  \left(
\tilde{\nabla}_{\nu}\phi\right)  -\frac{1}{2}g_{\mu\nu}\left(  \frac{1}%
{\sqrt{-g}}\left(  g^{\mu\nu}\sqrt{-g}\phi\right)  _{,\mu\nu}-g^{\mu\nu
}\left(  \tilde{\nabla}_{\mu}\phi\right)  \left(  \tilde{\nabla}_{\nu}%
\phi\right)  \right)  , \label{ww.04}%
\end{equation}
from which the Ricci scalar follows:%
\begin{equation}
\tilde{R}=R-\frac{3}{\sqrt{-g}}\left(  g^{\mu\nu}\sqrt{-g}\phi\right)
_{,\mu\nu}+\frac{3}{2}\left(  \tilde{\nabla}_{\mu}\phi\right)  \left(
\tilde{\nabla}_{\nu}\phi\right)  . \label{ww.05}%
\end{equation}

The simplest gravitational action Integral which can be defined in Weyl
integrable theory as an extension of the Einstein-Hilbert action has been
proposed to be,
\begin{equation}
S_{W}=\int dx^{4}\sqrt{-g}\left(  \tilde{R}+\xi\left(  \tilde{\nabla}_{\nu
}\left(  \tilde{\nabla}_{\mu}\phi\right)  \right)  g^{\mu\nu}\right)  ,
\label{ww.0a}%
\end{equation}
where $\xi$ is an arbitrary coupling constant,

Variation with respect to the metric tensor of (\ref{ww.0a}) provides the
field equations \cite{salim96},%
\begin{equation}
\tilde{G}_{\mu\nu}+\tilde{\nabla}_{\nu}\left(  \tilde{\nabla}_{\mu}%
\phi\right)  -\left(  2\xi-1\right)  \left(  \tilde{\nabla}_{\mu}\phi\right)
\left(  \tilde{\nabla}_{\nu}\phi\right)  +\xi g_{\mu\nu}g^{\kappa\lambda
}\left(  \tilde{\nabla}_{\kappa}\phi\right)  \left(  \tilde{\nabla}_{\lambda
}\phi\right)  =0, \label{ww.0b}%
\end{equation}
while variation with respect to the field $\phi$ gives
\begin{equation}
\left(  \tilde{\nabla}_{\nu}\left(  \tilde{\nabla}_{\mu}\phi\right)  \right)
g^{\mu\nu}+2g^{\mu\nu}\left(  \tilde{\nabla}_{\mu}\phi\right)  \left(
\tilde{\nabla}_{\nu}\phi\right)  =0 \label{ww.0c}%
\end{equation}
where $\tilde{G}_{\mu\nu}$ is the Einstein tensor in Weyl theory, that is,
$\tilde{G}_{\mu\nu}=\tilde{R}_{\mu\nu}-\frac{1}{2}\tilde{R}g_{\mu\nu}.$

However, we always write the field equations by using the geometric
definitions of Riemannian geometry, which means that equations (\ref{ww.0b})
are written in a similar form
\begin{equation}
G_{\mu\nu}-\zeta\left(  \phi_{,\mu}\phi_{,\nu}-\frac{1}{2}g_{\mu\nu}%
\phi^{,\kappa}\phi_{,\kappa}\right)  =0,
\end{equation}
where $2\zeta\equiv4\xi-3$, while for the scalar field $\phi$, the
second-order differential equation (\ref{ww.0c}) is the usual Klein-Gordon
equation $g^{\mu\nu}\nabla_{\nu}\nabla_{\mu}\phi=0.$

The origin of the scalar field $\phi$ is geometrical and it is related to the
nature of the affine connection, $\tilde{\Gamma}_{\mu\nu}^{\kappa}$.

\section{Einstein-Aether theory in Weyl integrable gravity}

\label{sec2a}

We consider the contribution of a timelike vector field, the so called aether
field $u^{\mu}$, in the gravitational action integral. In particular, we
assume the gravitational action Integral%
\begin{equation}
S_{AE}=S_{W}+S_{AE}, \label{ae.01}%
\end{equation}
where $S_{W}$ is the action integral (\ref{ww.0a}), and $S_{AE}$ is the action
integral which is defined by the aether field, that is,
\cite{Carruthers:2010ii}
\begin{equation}
S_{AE}=\int d^{4}x\sqrt{-g}\left(  K^{\alpha\beta\mu\nu}\tilde{\nabla}%
_{\alpha}u_{\mu}\tilde{\nabla}_{\beta}u_{\nu}+\lambda\left(  u^{c}%
u_{c}+1\right)  \right)  , \label{ae.01a}%
\end{equation}
where the Lagrange multiplier $\lambda~$ensures the unitarity of the aether
field, i.e. $u^{\mu}u_{\mu}=-1,$ and the fourth-rank tensor~$K^{\alpha\beta
\mu\nu}$ is expressed as follows:
\begin{equation}
K^{\alpha\beta\mu\nu}\equiv c_{1}g^{\alpha\beta}g^{\mu\nu}+c_{2}g^{\alpha\mu
}g^{\beta\nu}+c_{3}g^{\alpha\nu}g^{\beta\mu}+c_{4}g^{\mu\nu}u^{\alpha}%
u^{\beta}. \label{ae.02}%
\end{equation}
Parameters $c_{1},~c_{2},~c_{3}$ and $c_{4}$ are dimensionless constants and
define the coupling between the \ae ther field and gravity.

Variation with respect to the metric tensor in (\ref{ae.01}) provides the
gravitational field equations:%
\begin{equation}
\tilde{G}_{\mu\nu}+\tilde{\nabla}_{\nu}\left(  \tilde{\nabla}_{\mu}%
\phi\right)  -\left(  2\xi-1\right)  \left(  \tilde{\nabla}_{\mu}\phi\right)
\left(  \tilde{\nabla}_{\nu}\phi\right)  +\xi g_{\mu\nu}g^{\kappa\lambda
}\left(  \tilde{\nabla}_{\kappa}\phi\right)  \left(  \tilde{\nabla}_{\lambda
}\phi\right)  ={T_{ab}^{\ae },} \label{ae.022}%
\end{equation}
where
\begin{align}
{T_{ab}^{\ae }}  &  =2c_{1}(\tilde{\nabla}_{a}u^{c}\tilde{\nabla}_{b}%
u_{c}-\tilde{\nabla}_{c}u_{a}\tilde{\nabla}_{d}u_{b}g^{cd})+2\lambda
u_{a}u_{b}+g_{ab}\Phi_{u}\nonumber\\
&  -2[\tilde{\nabla}_{c}(u_{(a}J^{c}{}_{b)})+\tilde{\nabla}_{c}(u^{c}%
J_{(ab)})-\tilde{\nabla}_{c}(u_{(a}J_{b)}{}^{c})]-2c_{4}\left(  \tilde{\nabla
}u_{a}u^{c}\right)  \left(  \tilde{\nabla}u_{b}u^{d}\right)  , \label{ae.023}%
\end{align}
and%
\[
{{J}}{{^{a}}_{m}}=-{{K^{ab}}_{mn}\tilde{\nabla}u}^{n}~,~\Phi_{u}=-K^{ab}%
{}_{cd}\tilde{\nabla}_{a}u^{c}\tilde{\nabla}_{b}u^{d}\,,
\]
while the Lagrange multiplier $\lambda$ is defined by the constraint equation
\begin{equation}
c_{4}g^{\mu\nu}u^{\alpha}{\tilde{\nabla}}_{\beta}u_{\nu}{\tilde{\nabla}%
}_{\alpha}u_{\mu}g^{\kappa\beta}-c_{4}g^{\mu\kappa}g^{\alpha\lambda}%
{\tilde{\nabla}}_{\beta}u_{\lambda}u^{\beta}{\tilde{\nabla}}_{\alpha}u_{\mu
}-c_{4}g^{\mu\kappa}u^{\alpha}{\tilde{\nabla}}_{\beta}u^{\beta}{\tilde{\nabla
}}_{\alpha}u_{\mu}-K^{\alpha\beta\mu\kappa}{\tilde{\nabla}}_{\beta}%
{\tilde{\nabla}}_{\alpha}u_{\mu}-\lambda u^{\kappa}=0, \label{ae.04}%
\end{equation}
and the field $\phi$ satisfies equation (\ref{ww.0c}).

Using the kinematic quantities for the timelike vector field, the expansion
rate $\tilde{\theta},$ the shear $\tilde{\sigma}_{\mu\nu}$, the vorticity
$\tilde{\omega}_{\mu\nu}$, and the acceleration $\tilde{\alpha}^{\mu}$,
$\ $the action integral (\ref{ae.01a}) becomes \cite{DJ},%
\begin{equation}
S_{EA}=\int\sqrt{-g}dx^{4}\left(  \frac{c_{\theta}}{3}\tilde{\theta}%
^{2}+c_{\sigma}\tilde{\sigma}^{2}+c_{\omega}\tilde{\omega}^{2}+c_{\alpha
}\tilde{\alpha}^{2}\right)  , \label{ae.02a}%
\end{equation}
where the new parameters $c_{\theta},~c_{\sigma},~c_{\omega},~c_{a}$ are
functions of $c_{1},~c_{2},~c_{3}$ and $c_{4}$, i.e.
\begin{equation}
c_{\theta}=\left(  c_{1}+3c_{2}+c_{3}\right)  ~,\ c_{\sigma}=c_{1}%
+c_{3}~,\ c_{\omega}=c_{1}-c_{3}~,\ c_{a}=c_{4}-c_{1}, \label{ae.10}%
\end{equation}
and $\tilde{\sigma}^{2}=\tilde{\sigma}_{\mu\nu}\tilde{\sigma}^{\mu\nu}%
,~\tilde{\omega}^{2}=\tilde{\omega}_{\mu\nu}\tilde{\omega}^{\mu\nu}$, and
$\tilde{\alpha}^{2}=\tilde{\alpha}_{\mu}\tilde{\alpha}^{\mu}$.

\subsection{FLRW background spacetime}

For the homogeneous and isotropic FLRW spacetime with zero spatial curvature%

\begin{equation}
ds^{2}=-dt^{2}+a^{2}\left(  t\right)  \left(  dx^{2}+dy^{2}+dz^{2}\right)
,\label{fr0a}%
\end{equation}
and for the aether field $u^{\mu}=\delta_{t}^{\mu}$ we calculate
\begin{equation}
\tilde{\sigma}^{2}=0~,~\tilde{\omega}^{2}=0\text{ and }\tilde{\alpha}%
^{2}=0,\label{fr.0b}%
\end{equation}
while the energy-momentum tensor ${T_{ab}^{\ae }}$ can be written in the
simplest form%
\begin{equation}
{T_{ab}^{\ae }=}\rho^{\text{\ae \ }}u_{\mu}u_{\nu}+p^{\text{\ae }}h_{\mu\nu}%
\end{equation}
where $\rho^{\text{\ae \ }}=-\frac{c_{\theta}}{3}\tilde{\theta}^{2}$,
$~p^{\text{\ae \ }}=\frac{c_{\theta}}{3}\left(  2\tilde{\theta}_{,t}%
+\tilde{\theta}^{2}\right)  $ and $h_{\mu\nu}$ is the projection tensor
$h_{\mu\nu}=g_{\mu\nu}+u_{\mu}u_{\nu}$.

Therefore, for the line element (\ref{fr0a}), the gravitational field
equations are%
\begin{equation}
\frac{\theta^{2}}{3}-\rho_{\phi}-\rho^{\text{\ae \ }}=0 \label{ff.01}%
\end{equation}%
\begin{equation}
\dot{\theta}+\frac{\theta^{2}}{3}+\frac{1}{2}\left(  \rho_{\phi}+3p_{\phi
}\right)  +\frac{1}{2}\left(  \rho^{\text{\ae \ }}+3p^{\text{\ae }}\right)  =0
\label{ff.02}%
\end{equation}
where $\theta$ is the expansion rate of general relativity, that is,
$\theta=3\left(  \frac{\dot{a}}{a}\right)  ;$ and $\rho_{\phi}=\frac{\zeta}%
{2}\dot{\phi}^{2}$,~$p_{\phi}=\frac{\zeta}{2}\dot{\phi}^{2}$ are the energy
density and pressure for the field $\phi$.

The field equations (\ref{ff.01})-(\ref{ff.02}) can be written in an
equivalent way, as follows:%
\begin{equation}
\left(  1+c_{\theta}\right)  \frac{\theta^{2}}{3}-\frac{2}{3}c_{\theta}%
\theta\dot{\phi}-\left(  \frac{\zeta}{2}-\frac{c_{\theta}}{3}\right)
\dot{\phi}^{2}=0, \label{ff.04}%
\end{equation}%
\begin{equation}
\frac{\left(  1+c_{\theta}\right)  }{3}\dot{\theta}+\left(  1+c_{\theta
}\right)  \theta^{2}-\frac{2}{3}c_{\theta}\theta\dot{\phi}+\left(
\frac{c_{\theta}}{3}+\zeta\right)  \dot{\phi}^{2}-c_{\theta}\ddot{\phi}=0,
\label{ff.05}%
\end{equation}
while for the field $\phi$, we have%
\begin{equation}
\left(  2c_{\theta}-3\left(  1+c_{\theta}\right)  \zeta\right)  \ddot{\phi
}+3\zeta c_{\theta}\dot{\phi}^{2}-3\left(  1+c_{\theta}\right)  \zeta
\theta\dot{\phi}=0. \label{ff.06}%
\end{equation}

From (\ref{ff.05}) and (\ref{ff.06}, we find%
\begin{equation}
\frac{\dot{\theta}}{\theta^{2}}=-\frac{1}{3}-\frac{c_{\theta}}{3\left(
1+c_{\theta}\right)  \left(  3\zeta\left(  1+c_{\theta}\right)  -2c_{\theta
}\right)  }\left(  \frac{\dot{\phi}}{\theta}\right)  +\frac{\left(
2c_{\theta}-3\zeta\right)  \left(  c_{\theta}+3\zeta\left(  1+c_{\theta
}\right)  \right)  }{3\left(  1+c_{\theta}\right)  \left(  3\zeta\left(
1+c_{\theta}\right)  -2c_{\theta}\right)  }\left(  \frac{\dot{\phi}}{\theta
}\right)  ^{2},
\end{equation}
from which we infer that the parameter for the equation of state for the
effective fluid is $w_{eff}=-1-2\frac{\dot{\theta}}{\theta^{2}}.$

We have introduced a scalar field in Einstein-Aether theory by using the Weyl
geometry. Other attempts to introduce a scalar field in Einstein-Aether theory
have been proposed before by Kanno and Soda in \cite{Kanno:2006ty} and latter
by Donnelly and Jacobson in \cite{DJ}. In both attempts, the scalar field
interacts with the aether field. In \cite{Kanno:2006ty}, the interaction has
been proposed to be in the coupling parameters $c_{I}$ of the aether field. A
more general consideration has been proposed in\ \cite{DJ}, where an arbitrary
potential for the scalar field has been introduced in the action integral such
that kinematic quantities of the aether field are included. Our approach is
totally geometric and the interaction between the scalar field and the aether
field is introduced by the covariant derivative which defines the kinematic
quantities and the affine connection $\tilde{\Gamma}_{\mu\nu}^{\kappa}$ of the
Weyl geometry.

\subsubsection{Exact solution}

We select the new dimensionless variable $x\left(  t\right)  =\frac{2c_{\phi}%
}{1+c_{\phi}}\frac{\dot{\phi}}{\theta}$, to write the field equations as,%
\begin{align}
-12c_{\theta}^{2}\left(  3\zeta+c_{\theta}\left(  3\zeta-2\right)  \right)
\frac{dx}{d\ln a}  &  =x\left(  8c_{\theta}^{2}\left(  c_{\theta}%
+3\zeta\left(  1+c_{\theta}\right)  \right)  -8c_{\theta}^{2}\left(
c_{\theta}+3\zeta\left(  1+c\theta\right)  \right)  x\right) \nonumber\\
&  +x^{3}\left(  1+c_{\theta}\right)  \left(  3c_{\theta}\left(
1-3\zeta\right)  \zeta-9\zeta^{2}+\left(  2+6\zeta\right)  c_{\theta}%
^{2}\right)  ,
\end{align}
while the constraint equation becomes,%
\begin{equation}
1+\left(  \frac{\left(  1+c_{\theta}\right)  \left(  2c_{\theta}%
-3\zeta\right)  }{8c_{\theta}}x-1\right)  x=0.
\end{equation}

From the latter algebraic equation we get,
\begin{equation}
\mathbf{x^{\pm}}=\frac{2c_{\theta}}{1+c_{\theta}}\frac{2c_{\theta}\pm
\sqrt{6\zeta+c_{\theta}\left(  6\zeta-4\right)  }}{2c_{\theta}-3\zeta
},~2c_{\theta}-3\zeta\neq0.
\end{equation}
The points are real when $6\zeta+c_{\theta}\left(  6\zeta-4\right)  \geq
0\,.$These two points provide ideal-gas exact solutions with an equation of
state parameter,%
\begin{equation}
w_{eff}\left(  \mathbf{x^{\pm}}\left(  c_{\theta},\zeta\right)  \right)
=-1-2\frac{\dot{\theta}}{\theta^{2}}=\frac{9\zeta^{2}+c_{\theta}^{2}\left(
6\zeta-4\right)  +3\zeta c_{\theta}\left(  3\zeta+2\sqrt{6\left(  1+c_{\theta
}\right)  \zeta-4c_{\theta}}\right)  }{\left(  2c_{\theta}-3\zeta\right)
\left(  3\left(  1+c_{\theta}\right)  \zeta-2c_{\theta}\right)  }.
\end{equation}
In the special limit where \ $\zeta=0$, the exact solution is that of a stiff
fluid, that is, $w_{eff}\left(  \mathbf{x^{\pm}}\left(  c_{\theta},0\right)
\right)  =1$.

\begin{figure}[ptb]
\centering\includegraphics[width=0.7\textwidth]{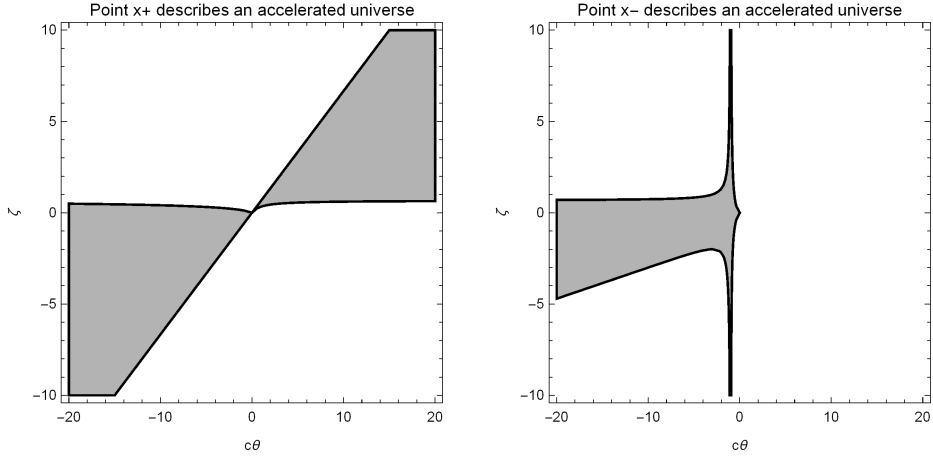} \caption{Region
plots in the space of variables $\left\{  c_{\theta},\zeta\right\}  $. The
exact solutions at the critical points $\mathbf{x^{+}}$ (left fig.) and
$\mathbf{x^{-}}$ (right fig.) describe accelerated universes. }%
\label{fig0}%
\end{figure}

In Fig. \ref{fig0} we present the region of the variables $\left\{  c_{\theta
},\zeta\right\}  $ in which the exact solutions at the critical points$~x^{\pm
}$ describe accelerated universes, that is, $w_{eff}\left(  \mathbf{x^{\pm}%
}\left(  c_{\theta},\zeta\right)  \right)  <-\frac{1}{3}$. 

\section{In the presence of matter}

\label{sec4}

If we consider now the existence of a pressureless fluid source, then the
field equations become,
\begin{equation}
G_{\mu\nu}-\zeta\left(  \phi_{,\mu}\phi_{,\nu}-\frac{1}{2}g_{\mu\nu}%
\phi^{,\kappa}\phi_{,\kappa}\right)  ={T_{ab}^{\ae }+}T_{\mu\nu}^{\left(
m\right)  }, \label{ww.12}%
\end{equation}
where~$T_{\mu\nu}^{\left(  m\right)  }=e^{-\frac{\phi}{2}}\rho_{m}u_{\mu
}u_{\nu}$,$~$while the conservation equation for the matter field reads
$\tilde{\nabla}_{\nu}T^{\left(  m\right)  \mu\nu}=0$.

For the spatially-flat FLRW line element (\ref{fr0a}), the conservation
equation for the dust fluid reads,%
\begin{equation}
\dot{\rho}_{m}+\left(  \theta-\dot{\phi}\right)  \rho_{m}=0, \label{ww.13}%
\end{equation}
while the rest of the field equations become%
\begin{equation}
\left(  1+c_{\theta}\right)  \frac{\theta^{2}}{3}-\frac{2}{3}c_{\theta}%
\theta\dot{\phi}-\left(  \frac{\zeta}{2}-\frac{c_{\theta}}{3}\right)
\dot{\phi}^{2}-e^{-\frac{\phi}{2}}\rho_{m}=0, \label{ww.14}%
\end{equation}%
\begin{equation}
\frac{\left(  1+c_{\theta}\right)  }{3}\dot{\theta}+\left(  1+c_{\theta
}\right)  \theta^{2}-\frac{2}{3}c_{\theta}\theta\dot{\phi}+\left(
\frac{c_{\theta}}{3}+\zeta\right)  \dot{\phi}^{2}-c_{\theta}\ddot{\phi}%
+\frac{1}{2}e^{-\frac{\phi}{2}}\rho_{m}=0, \label{ww.15}%
\end{equation}
while the Klein-Gordon equation is modified as%
\begin{equation}
\left(  2c_{\theta}-3\left(  1+c_{\theta}\right)  \zeta\right)  \ddot{\phi
}+3\zeta c_{\theta}\dot{\phi}^{2}-3\left(  1+c_{\theta}\right)  \zeta
\theta\dot{\phi}+3\left(  1-c_{\theta}\right)  \rho_{m}e^{-\frac{\phi}{2}}=0.
\label{ww.16}%
\end{equation}

With the use of the dimensionless variables,
\begin{equation}
x\equiv\frac{2c_{\phi}}{1+c_{\phi}}\frac{\dot{\phi}}{\theta}\text{ and }%
\Omega_{m}\equiv\frac{3e^{-\frac{\phi}{2}}\rho_{m}}{\left(  1+c_{\theta
}\right)  \theta^{2}}, \label{ww.16a}%
\end{equation}

the field equations reduce to the following algebraic-differential system,%
\begin{equation}
\Omega_{m}=1+\left(  \frac{\left(  1+c_{\theta}\right)  \left(  2c_{\theta
}-3\zeta\right)  }{8c_{\theta}}x-1\right)  x, \label{ww.17}%
\end{equation}%
\begin{equation}
-16c_{\theta}^{2}\left(  3\left(  1+c_{\theta}\right)  \zeta-2c_{\theta
}\right)  \frac{dx}{d\ln a}=\left(  \left(  1+c_{\theta}\right)  \left(
2c_{\theta}-3\zeta\right)  x^{2}+8c_{\theta}^{2}\left(  1-x\right)  \right)
\left(  \left(  1+c_{\theta}\right)  \left(  c_{\theta}+3\zeta\right)
x+2\left(  1-c_{\theta}\right)  c_{\theta}\right)  . \label{ww.18}%
\end{equation}

The stationary points of equation (\ref{ww.18}) are,%
\[
\mathbf{x^{\pm}}=\frac{2c_{\theta}}{1+c_{\theta}}\frac{2c_{\theta}\pm
\sqrt{6\zeta+c_{\theta}\left(  6\zeta-4\right)  }}{2c_{\theta}-3\zeta}\text{
and }\mathbf{x^{0}}=\frac{2c_{\theta}\left(  c_{\theta}-1\right)  }{\left(
1+c_{\theta}\right)  \left(  c_{\theta}+3\zeta\right)  }.
\]

Points \ $\mathbf{x^{\pm}}$ are the critical points of the vacuum case,
$\Omega_{m}\left(  \mathbf{x^{\pm}}\right)  =0$, while point $\mathbf{x^{0}}$
provides
\begin{equation}
\Omega_{m}\left(  \mathbf{x^{0}}\right)  =\frac{c_{\theta}+9\zeta^{2}}{\left(
c_{\theta}+3\zeta\right)  ^{2}}-\frac{3\left(  1+c_{\theta}\left(  c_{\theta
}-10\right)  \right)  \zeta}{\left(  1+c_{\theta}\right)  \left(  c_{\theta
}+3\zeta\right)  ^{2}}. \label{ww.19}%
\end{equation}

\begin{figure}[ptb]
\centering\includegraphics[width=0.7\textwidth]{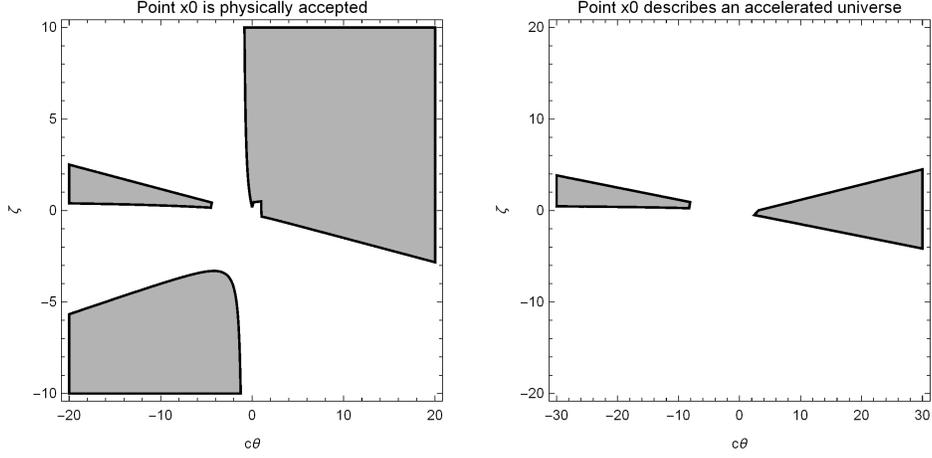} \caption{Region
plots in the space of variables $\left\{  c_{\theta},\zeta\right\}  $, where
the critical point $\mathbf{x^{0}}$ is physically accepted (left fig.) and the
exact solution at this point describes an accelerated universe (right fig.).}%
\label{fig1}%
\end{figure}

Point $\mathbf{x^{0}}$ is physically accepted when $0\leq\Omega_{m}\left(
\mathbf{x^{0}}\right)  \leq1$, as presented in Fig. \ref{fig1}

As far as the equation of state for the cosmological solution at point
$\mathbf{x^{0}}$ is concerned, it is calculated to be,
\begin{equation}
w_{eff}\left(  \mathbf{x^{0}}\left(  c_{\theta},\zeta\right)  \right)
=\frac{1-c_{\theta}}{2\left(  c_{\theta}+6\zeta\right)  }. \label{ww.20}%
\end{equation}
The point $\mathbf{x^{0}}~$describes an accelerated universe,~i.e. $w\left(
\mathbf{x^{0}}\left(  c_{\theta},\zeta\right)  \right)  <-\frac{1}{3}$, when
$\zeta,c_{\theta}$ take values in the range of Fig. \ref{fig1}.

As far as the stability of the stationary points is concerned, we determine
the eigenvalue $e\left(  \mathbf{x}\left(  c_{\theta},\zeta\right)  \right)  $
of equation (\ref{ww.18}). A stationary point is an attractor when~
$\operatorname{Re}\left(  e\left(  \mathbf{x}\left(  c_{\theta},\zeta\right)
\right)  \right)  <0\,$. In particular, we find that $e\left(  \mathbf{x^{0}%
}\left(  c_{\theta},\zeta\right)  \right)  =\frac{\left(  1+c_{\theta}\right)
\left(  c_{\theta}+3\zeta\right)  }{6\left(  1+c_{\theta}\right)
\zeta-4c_{\theta}}\Omega_{m}\left(  \mathbf{x^{0}}\right)  $, while, for the
other two points, it follows
\begin{equation}
e\left(  \mathbf{x^{\pm}}\left(  c_{\theta},\zeta\right)  \right)
=\frac{18\zeta^{2}+6c_{\theta}\left(  3\zeta-1\right)  +c_{\theta}^{2}\left(
6\zeta-4\right)  \mp3\left(  \zeta-c_{\theta}\left(  2+9\zeta\right)  \right)
\sqrt{6\left(  1+c_{\theta}\right)  \zeta-4c_{\theta}}}{2\left(  2c_{\theta
}-3\zeta\right)  \left(  3\left(  1+c_{\theta}\right)  \zeta-2c_{\theta
}\right)  }.
\end{equation}

\begin{figure}[ptb]
\centering\includegraphics[width=1\textwidth]{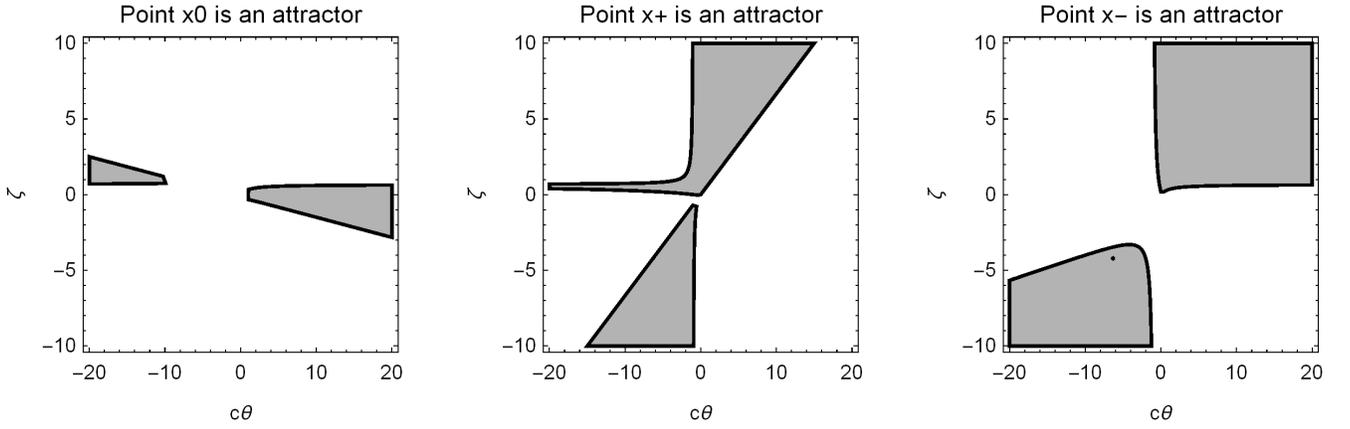} \caption{Region
plots in the space of variables $\left\{  c_{\theta},\zeta\right\}  $, where
the stationary points: $\mathbf{x^{0}}$ (left fig.), $\mathbf{x^{+}}$ (Middle
Fig.) and $\mathbf{x^{-}}$ (right fig.) are attractors for the cosmological
model with a dust fluid source.}%
\label{fig2}%
\end{figure}

In Fig. \ref{fig2}, we present the regions in the space of variables,~$\left(
c_{\theta},\zeta\right)  $, where the stationary points are stable.

In the limit where $\zeta=0$, for point $\mathbf{x^{0}}$ it follows
$\Omega_{m}\left(  \mathbf{x^{0}}\right)  =\frac{1}{c_{\theta}}$ and
$w_{eff}\left(  \mathbf{x^{0}}\left(  c_{\theta},2\right)  \right)  =-\frac
{1}{2}\left(  1-c_{\theta}\right)  ,$ $\ $hence it follows that point
$\mathbf{x^{0}}$ is physically accepted when $c_{\theta}\geq1$ while for
$c_{\theta}>3$ the exact solution describes an accelerated universe. That is a
very interesting result because $\Omega_{m}\left(  \mathbf{x^{0}}\right)  $
becomes zero only for very large values of the coupling constant $c_{\theta}$,
while $e\left(  \mathbf{x^{0}}\left(  c_{\theta},0\right)  \right)  $ is found
to be always negative for $c_{\theta}>1$ which means that $\mathbf{x^{0}}$ is
a future attractor.

\section{Evolution of a scalar field under potential $V\left(  \phi\right)  $
in vacuum}

\label{sec5}

In the presence of a potential term $V\left(  \phi\right)  $, the field
equations (\ref{ff.01})-(\ref{ff.02}) become,%
\begin{equation}
\left(  1+c_{\theta}\right)  \frac{\theta^{2}}{3}-\frac{2}{3}c_{\theta}%
\theta\dot{\phi}-\left(  \frac{\zeta}{2}-\frac{c_{\theta}}{3}\right)
\dot{\phi}^{2}-V\left(  \phi\right)  =0, \label{ff.07}%
\end{equation}%
\begin{equation}
\frac{\left(  1+c_{\theta}\right)  }{3}\dot{\theta}+\left(  1+c_{\theta
}\right)  \theta^{2}-\frac{2}{3}c_{\theta}\theta\dot{\phi}+\left(
\frac{c_{\theta}}{3}+\zeta\right)  \dot{\phi}^{2}-c_{\theta}\ddot{\phi
}-V\left(  \phi\right)  =0, \label{ff.08}%
\end{equation}
while the field $\phi$ obeys,%
\begin{equation}
\left(  2c_{\theta}-3\left(  1+c_{\theta}\right)  \zeta\right)  \ddot{\phi
}+3\zeta c_{\theta}\dot{\phi}^{2}-3\left(  1+c_{\theta}\right)  \zeta
\theta\dot{\phi}-3\left(  1+c_{\theta}\right)  V_{,\phi}=0. \label{ff.09}%
\end{equation}

We define the new dimensionless variables, $x,$ from (\ref{ww.16a}) and
$y^{2}\equiv\frac{3V\left(  \phi\right)  }{\left(  1+c_{\theta}\right)
\theta^{2}}$, and the field equations become,%
\begin{equation}
1+\left(  \frac{\left(  1+c_{\theta}\right)  \left(  2c_{\theta}%
-3\zeta\right)  }{8c_{\theta}}x-1\right)  x-y^{2}=0,
\end{equation}%
\begin{align}
-24c_{\theta}^{2}\left(  3\left(  1+c_{\theta}\right)  \zeta-2c_{\theta
}\right)  \frac{dx}{d\ln a}  &  =\left(  \left(  1+c_{\theta}\right)  \left(
2c_{\theta}-3\zeta\right)  x^{2}+8c_{\theta}^{2}\left(  1-x\right)  \right)
\times\nonumber\\
&  \times\left(  x\left(  2c_{\theta}+3\left(  1+c_{\theta}\right)  \left(
2\zeta+\lambda c_{\theta}\right)  \right)  -6\lambda c_{\theta}\left(
1+c_{\theta}\right)  \right)  ,
\end{align}
where $\lambda=-\frac{V_{\phi}}{V}$. For the scalar field, we assume the
exponential potential $V\left(  \phi\right)  =V_{0}e^{-\lambda\phi}$ where
$\lambda=$constant. The exponential potential is of special interest,
mathematically and physically. In terms of mathematics, it reduces the
dimension of the dynamical system, while, in terms of physics ,when
$\lambda=2$, it includes the case of the cosmological constant term in Weyl
integrable theory \cite{salim96}.

The latter dynamical system admits the stationary points $\mathbf{x^{\pm}}$
and $\mathbf{x^{1}}=\frac{2c_{\theta}\lambda}{3\zeta+c_{\theta}\lambda}$.
\ The new stationary point $\mathbf{x^{1}}$ describes a universe where the
scalar field potential also contributes to the effective cosmological fluid,
while the equation of state parameter $w_{eff}\left(  \mathbf{x^{1}}\right)  $
for the effective fluid at the stationary point is found to be,
\begin{equation}
w_{eff}\left(  \mathbf{x^{1}}\left(  c_{\theta},\zeta,\lambda\right)  \right)
=\frac{\lambda\left(  c_{\theta}\left(  \lambda-1\right)  +\lambda\right)
-3\zeta}{3\zeta+c_{\theta}\lambda}.
\end{equation}
The corresponding eigenvalues are derived%
\begin{align}
2(3(\text{$c_{\theta}$}+1)\zeta-2\text{$c_{\theta}$})(\text{$c_{\theta}$%
}\lambda+3\zeta)^{2}e\left(  \mathbf{x^{1}}\left(  c_{\theta},\zeta
,\lambda\right)  \right)   &  =-12\zeta^{2}(3(\text{$c_{\theta}$}%
+1)\zeta+\text{$c_{\theta}$})+\text{$c_{\theta}$}(\text{$c_{\theta}$%
}+1)\lambda^{3}(3(\text{$c_{\theta}$}+1)\zeta-2\text{$c_{\theta}$%
})+\nonumber\\
&  +2\text{$c_{\theta}$}\lambda^{2}(3(\text{$c_{\theta}$}+1)\zeta
-2\text{$c_{\theta}$})-2\text{$c_{\theta}$}\zeta\lambda(15(\text{$c_{\theta}$%
}+1)\zeta-4\text{$c_{\theta}$})
\end{align}%
\begin{align}
E_{1}e\left(  \mathbf{x^{+}}\left(  c_{\theta},\zeta,\lambda\right)  \right)
&  =\left(  8\text{$c_{\theta}$}^{2}-4\text{$c_{\theta}$}\left(
\sqrt{6(\text{$c_{\theta}$}+1)\zeta-4\text{$c_{\theta}$}}+2\right)  \right)
\times\nonumber\\
&  \times\left(
\begin{array}
[c]{c}%
3\text{$c_{\theta}$}(\text{$c_{\theta}$}+1)^{2}\lambda(2\text{$c_{\theta}$%
}-3\zeta)-\text{$c_{\theta}$}\left(  \sqrt{6(\text{$c_{\theta}$}%
+1)\zeta-4\text{$c_{\theta}$}}+2\right)  \times\\
(6(\text{$c_{\theta}$}+1)\zeta+3(\text{$c_{\theta}$}+1)\text{$c_{\theta}$%
}\lambda+2\text{$c_{\theta}$})
\end{array}
\right)  +\nonumber\\
&  -8(\text{$c_{\theta}$}-1)\text{$c_{\theta}$}^{2}\left(  \sqrt
{6(\text{$c_{\theta}$}+1)\zeta-4\text{$c_{\theta}$}}-\text{$c_{\theta}$%
}+1\right)  (6(\text{$c_{\theta}$}+1)\zeta+3(\text{$c_{\theta}$}%
+1)\text{$c_{\theta}$}\lambda+2\text{$c_{\theta}$}).
\end{align}%
\begin{align}
E_{1}e\left(  \mathbf{x^{-}}\left(  c_{\theta},\zeta,\lambda\right)  \right)
&  =8(\text{$c_{\theta}$}-1)\text{$c_{\theta}$}^{2}\left(  \sqrt
{6(\text{$c_{\theta}$}+1)\zeta-4\text{$c_{\theta}$}}+\text{$c_{\theta}$%
}-1\right)  \times\nonumber\\
&  \times(6(\text{$c_{\theta}$}+1)\zeta+3(\text{$c_{\theta}$}%
+1)\text{$c_{\theta}$}\lambda+2\text{$c_{\theta}$})+4\text{$c_{\theta}$%
}\left(  \sqrt{6(\text{$c_{\theta}$}+1)\zeta-4\text{$c_{\theta}$}%
}+2\text{$c_{\theta}$}-2\right)  +\nonumber\\
&  +\left(
\begin{array}
[c]{c}%
3\text{$c_{\theta}$}(\text{$c_{\theta}$}+1)^{2}\lambda(2\text{$c_{\theta}$%
}-3\zeta)+\text{$c_{\theta}$}\left(  \sqrt{6(\text{$c_{\theta}$}%
+1)\zeta-4\text{$c_{\theta}$}}-2\right)  \times\\
\times(6(\text{$c_{\theta}$}+1)\zeta+3(\text{$c_{\theta}$}+1)\text{$c_{\theta
}$}\lambda+2\text{$c_{\theta}$})
\end{array}
\right)  .
\end{align}
where $E_{1}=12c_{\theta}$$^{2}(c_{\theta}$$+1)(2c_{\theta}$$-3\zeta
)(2c_{\theta}$$-3(c_{\theta}$$+1)\zeta)$.

\begin{figure}[ptb]
\centering\includegraphics[width=1\textwidth]{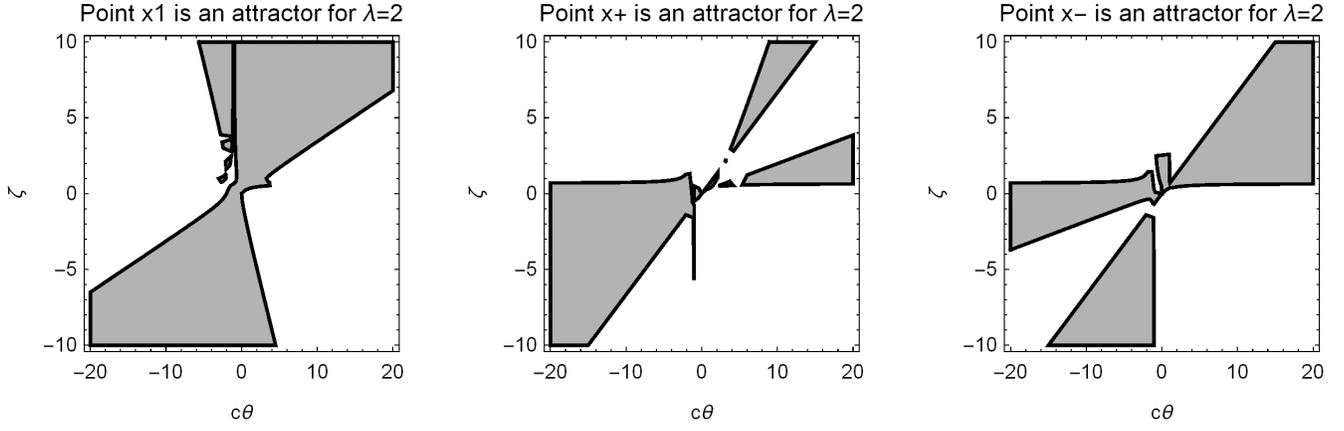} \caption{Region
plots in the space of variables $\left\{  c_{\theta},\zeta\right\}  $, where
the stationary points: $\mathbf{x^{1}}$ (left fig.), $\mathbf{x^{+}}$ (middle
fig.) and $\mathbf{x^{-}}$ (right fig.) are attractors for the dynamical
system with an exponential potential and $\lambda=2.$}%
\label{fig3}%
\end{figure}

\begin{figure}[ptb]
\centering\includegraphics[width=0.3\textwidth]{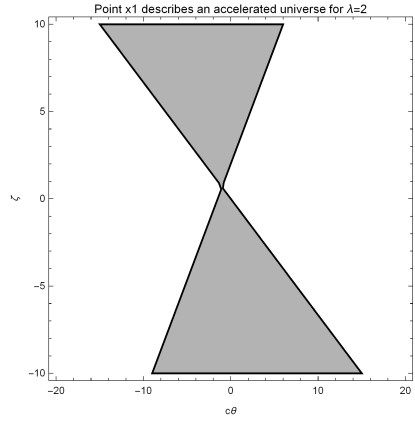} \caption{Region
plots in the space of variables $\left\{  c_{\theta},\zeta\right\}  $, where
the stationary point $\mathbf{x^{1}}$ describes and accelerated universe when
$\lambda=2$, that is $w_{eff}\left(  \mathbf{x^{1}}\left(  c_{\theta}%
,\zeta,2\right)  \right)  <-\frac{1}{3}$. }%
\label{fig4}%
\end{figure}

We focus on the special value where $\lambda=2~$\ which, as we described
before, corresponds to the case of the cosmological constant. In Fig.
\ref{fig3}, we present the regions in the space of the free parameters
$\left(  c_{\theta},\zeta\right)  ,$ where the three stationary points
$\mathbf{x^{1}},~\mathbf{x^{\pm}}$ are attractors, while in Fig. \ref{fig4} we
present the region of the free parameters $\left(  c_{\theta},\zeta\right)  $
where $w_{eff}\left(  \mathbf{x^{1}}\left(  c_{\theta},\zeta,\lambda\right)
\right)  <-\frac{1}{3}$ for $\lambda=2$.

Recall that for point $\mathbf{x^{1}}$ is physically acceptable as long as
$y^{2}\left(  \mathbf{x^{1}}\right)  \geq0,$ where for $\lambda=2$ we find the
range~$\left\{  c_{\theta}\leq-1,~\zeta\neq-\frac{2}{3}c_{\theta}\right\}
$,~$\left\{  -1<c_{\theta}<0,~\zeta<\frac{2}{3}c_{\theta}~,~\zeta>\frac{2}%
{3}\right\}  $, $\left\{  0<c_{\theta}\leq1~,~\zeta\neq-\frac{2}{3}c_{\theta
}\right\}  $,~$\left\{  c_{\theta}>1~,~\zeta\neq-\frac{2}{3}c_{\theta}%
,~\zeta<\frac{2}{3}~,~\zeta>\frac{2}{3}c_{\theta}~\right\}  $.

In the case where $\zeta=0$, point $\mathbf{x^{1}}$ is physically accepted
when $c_{\theta}\geq1$, while $w_{eff}\left(  \mathbf{x^{1}}\left(  c_{\theta
},\zeta,\lambda\right)  \right)  =-1+\lambda\left(  1+\frac{1}{c_{\theta}%
}\right)  $, and the point is an attractor when $\lambda\leq-1$ and $\left\{
-1<\lambda<0,~c_{\theta}>-\left(  1+\frac{2}{\lambda}\right)  \right\}  $.

\section{Evolution of a scalar field with potential $V\left(  \phi\right)  $
and matter}

In this example the Friedmann equation, Raychaudhuri, and Klein-Gordon
equations are, respectively,
\begin{align}
&  (2c_{\theta}-3\zeta)\dot{\phi}^{2}-4c_{\theta}\theta\dot{\phi}+2(c_{\theta
}+1)\theta^{2}-6\rho_{m}e^{-\frac{\phi}{2}}-6V(\phi)=0,\\
&  6c_{\theta}\dot{\theta}-4c_{\theta}\theta\dot{\phi}+2(c_{\theta}%
+1)\theta^{2}-6c_{\theta}\ddot{\phi}+2c_{\theta}\dot{\phi}^{2}+6\zeta\dot
{\phi}^{2}+6\dot{\theta}+3\rho_{m}e^{-\frac{\phi}{2}}-6V(\phi)=0,\\
&  3(c_{\theta}-1)\rho_{m}-2e^{\frac{\phi}{2}}\left(  3\zeta\dot{\phi}\left(
c_{\theta}\theta-c_{\theta}\dot{\phi}+\theta\right)  +(3(c_{\theta}%
+1)\zeta-2c_{\theta})\ddot{\phi}+3(c_{\theta}+1)V^{\prime}(\phi)\right)  =0,
\end{align}
plus the conservation equation \eqref{ww.13}.

Defining $x\equiv\frac{2c_{\phi}}{1+c_{\phi}}\frac{\dot{\phi}}{\theta}$ and
$\Omega_{m}\equiv\frac{3e^{-\frac{\phi}{2}}\rho_{m}}{\left(  1+c_{\theta
}\right)  \theta^{2}},$ $y^{2}\equiv\frac{3V(\phi)}{(c_{\theta}+1)\theta^{2}}%
$, we obtain the dynamical system:
\begin{align}
&  x^{\prime}=-\frac{(c_{\theta}+1)^{2}(2c_{\theta}-3\zeta)(c_{\theta}%
+3\zeta)x^{3}}{16c_{\theta}^{2}(3(c_{\theta}+1)\zeta-2c_{\theta})}+x\left(
\frac{-3c_{\theta}^{2}-3c_{\theta}\zeta+c_{\theta}-3\zeta}{6(c_{\theta
}+1)\zeta-4c_{\theta}}+\frac{(c_{\theta}+1)y^{2}(c_{\theta}(2\lambda
-1)+3\zeta)}{4c_{\theta}-6(c_{\theta}+1)\zeta}\right) \nonumber\\
&  +\frac{(c_{\theta}-1)c_{\theta}}{3(c_{\theta}+1)\zeta-2c_{\theta}}%
+\frac{(c_{\theta}+1)(c_{\theta}(6c_{\theta}+9\zeta-2)+3\zeta)x^{2}%
}{8c_{\theta}(3(c_{\theta}+1)\zeta-2c_{\theta})}+\frac{c_{\theta}(2(c_{\theta
}+1)\lambda-c_{\theta}+1)y^{2}}{3(c_{\theta}+1)\zeta-2c_{\theta}}%
,\label{r01}\\
&  2y^{\prime}=-\frac{(c_{\theta}+1)^{2}(2c_{\theta}-3\zeta)(c_{\theta}%
+3\zeta)x^{2}y}{8c_{\theta}^{2}(3(c_{\theta}+1)\zeta-2c_{\theta})}+xy\left(
\frac{c_{\theta}(c_{\theta}+1)}{3(c_{\theta}+1)\zeta-2c_{\theta}}%
-\frac{(c_{\theta}+1)\lambda}{2c_{\theta}}\right) \nonumber\\
&  +\frac{(c_{\theta}+1)y^{3}(-2c_{\theta}\lambda+c_{\theta}-3\zeta
)}{3(c_{\theta}+1)\zeta-2c_{\theta}}-\frac{(c_{\theta}+1)(c_{\theta}-3\zeta
)y}{3(c_{\theta}+1)\zeta-2c_{\theta}}, \label{r02}%
\end{align}
where $\lambda=-\frac{V_{\phi}}{V}$ and for the scalar field potential we
assume the exponential potential $V\left(  \phi\right)  =V_{0}e^{-\lambda\phi
}$ with constant $\lambda$. The fractional energy density of matter is
\begin{equation}
\Omega_{m}=\frac{(c_{\theta}+1)(2c_{\theta}-3\zeta)x^{2}}{8c_{\theta}^{2}%
}-x-y^{2}+1.
\end{equation}

Therefore, the phase plane is defined by
\begin{equation}
\left\{  (x,y):0\leq x-\frac{(c_{\theta}+1)(2c_{\theta}-3\zeta)x^{2}%
}{8c_{\theta}^{2}}+y^{2}\leq1,y\geq0\right\}  .
\end{equation}

The stationary points of the dynamical system (\ref{r01}), (\ref{r02}) are the
points $\mathbf{x^{0}}$,~$\mathbf{x^{\pm}}$,~$\mathbf{x^{1}}$ and a new point
$\mathbf{x}^{2}$, with coordinates
\[
\mathbf{x}^{2}=\left(  \frac{4c_{\theta}}{\left(  1+2\lambda\right)  \left(
c_{\theta}+1\right)  },\frac{\sqrt{-2c_{\theta}\lambda+c_{\theta}%
+6\zeta+2\lambda+1}}{\sqrt{c_{\theta}+1}(2\lambda+1)}\right)  .
\]

\begin{figure}[ptb]
\centering\includegraphics[width=0.7\textwidth]{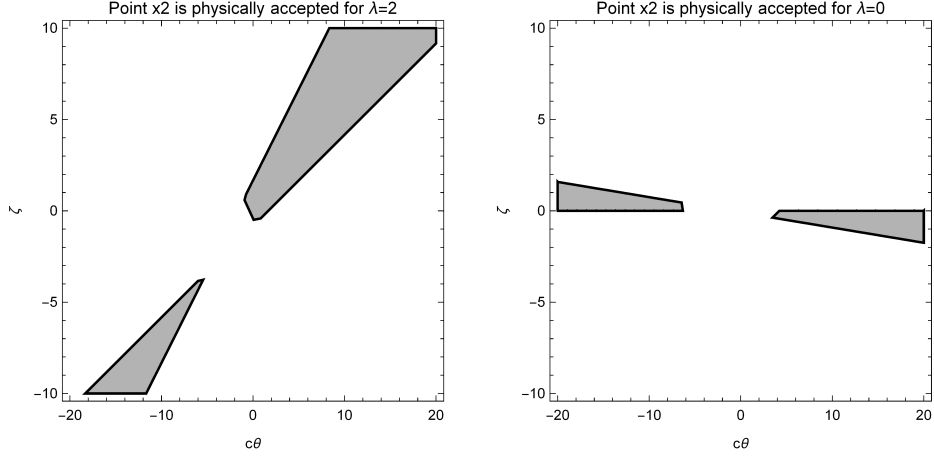} \caption{Region
plots in the space of variables $\left\{  c_{\theta},\zeta\right\}  $, where
the stationary point $\mathbf{x}^{2}$ is physically acceptable, for
$\lambda=2$ (left fig.) and $\lambda=0$ (right fig.).}%
\label{fig5}%
\end{figure}

For the exact solution at the point $\mathbf{x}^{2}$ we find $\Omega
_{m}\left(  \mathbf{x}^{2}\right)  =\frac{2\left(  \lambda\left(  1-c_{\theta
}+2\left(  1+c_{\theta}\right)  \lambda\right)  -6\zeta\right)  }{\left(
1+c_{\theta}\right)  \left(  1+2\lambda\right)  ^{2}}$ and
\begin{align}
3(\text{$c_{\theta}$}+1)(2\lambda+1)^{2}(3(\text{$c_{\theta}$}+1)\zeta
-2\text{$c_{\theta}$})w_{eff}\left(  \mathbf{x^{2}}\left(  c_{\theta}%
,\zeta,\lambda\right)  \right)   &  =36(\text{$c_{\theta}$}+1)\zeta
^{2}+4(\text{$c_{\theta}$}+1)\lambda^{2}(\text{$c_{\theta}$}(3\text{$c_{\theta
}$}-1)-3(\text{$c_{\theta}$}+1)\zeta)+\nonumber\\
&  -2\lambda(6(\text{$c_{\theta}$}+1)(\text{$c_{\theta}$}+2)\zeta
+\text{$c_{\theta}$}(\text{$c_{\theta}$}(3\text{$c_{\theta}$}%
-10)-5))+\nonumber\\
&  -3(\text{$c_{\theta}$}(15\text{$c_{\theta}$}+2)+3)\zeta+6\text{$c_{\theta}%
$}(\text{$c_{\theta}$}+1),
\end{align}
which means that the stationary point $\mathbf{x}^{2}$ describes a scaling
solution. The point is physically acceptable when $0\leq\Omega_{m}\left(
\mathbf{x}^{2}\right)  \leq1$. For the specific cases, $\lambda=0$ and
$\lambda=2,$ in Fig. \ref{fig5} we present the range of the parameters
$\left\{  c_{\theta},\zeta\right\}  $ for which the point is physically
acceptable. We continue with the stability analysis of the stationary points.

For point $\mathbf{x^{0}}$, the eigenvalues of the linearized system are
derived%
\begin{align*}
e_{1}\left(  \mathbf{x^{0}}\left(  c_{\theta},\zeta,\lambda\right)  \right)
&  =\frac{-18(c_{\theta}+1)\zeta^{2}+3((c_{\theta}-10)c_{\theta}%
+1)\zeta-2c_{\theta}(c_{\theta}+1)}{4(c_{\theta}+3\zeta)(3(c_{\theta}%
+1)\zeta-2c_{\theta})},\\
e_{2}\left(  \mathbf{x^{0}}\left(  c_{\theta},\zeta,\lambda\right)  \right)
&  =\frac{-2c_{\theta}\lambda+c_{\theta}+6\zeta+2\lambda+1}{4c_{\theta
}+12\zeta}.
\end{align*}
Similarly, for the rest of the points we find%
\begin{align*}
e_{1}\left(  \mathbf{x^{+}}\left(  c_{\theta},\zeta,\lambda\right)  \right)
&  =\sqrt{6(c_{\theta}+1)\zeta-4c_{\theta}}\left(  \frac{3}{3\zeta-2c_{\theta
}}-\frac{4}{3(c_{\theta}+1)\zeta-2c_{\theta}}\right)  +\frac{3c_{\theta}%
}{3\zeta-2c_{\theta}}+1,\\
e_{2}\left(  \mathbf{x^{+}}\left(  c_{\theta},\zeta,\lambda\right)  \right)
&  =\frac{\left(  -8c_{\theta}^{2}(c_{\theta}-3\zeta)-\frac{16c_{\theta}%
^{3}\left(  2c_{\theta}^{2}+c_{\theta}(2-3\zeta)\lambda-3\zeta\lambda\right)
}{\sqrt{6c_{\theta}^{2}(c_{\theta}+1)\zeta-4c_{\theta}^{3}}-2c_{\theta}^{2}%
}-\frac{16(c_{\theta}+1)c_{\theta}^{4}(2c_{\theta}-3\zeta)(c_{\theta}+3\zeta
)}{\left(  \sqrt{6c_{\theta}^{2}(c_{\theta}+1)\zeta-4c_{\theta}^{3}%
}-2c_{\theta}^{2}\right)  ^{2}}\right)  }{16c_{\theta}^{2}(3(c_{\theta
}+1)\zeta-2c_{\theta})(c_{\theta}+1)^{-1}},
\end{align*}%
\begin{align*}
e_{1}\left(  \mathbf{x^{-}}\left(  c_{\theta},\zeta,\lambda\right)  \right)
&  =\sqrt{6(c_{\theta}+1)\zeta-4c_{\theta}}\left(  \frac{3}{3\zeta-2c_{\theta
}}+\frac{4}{3(c_{\theta}+1)\zeta-2c_{\theta}}\right)  +\frac{3c_{\theta}%
}{3\zeta-2c_{\theta}}+1,\\
e_{2}\left(  \mathbf{x^{-}}\left(  c_{\theta},\zeta,\lambda\right)  \right)
&  =\frac{\left(  -8c_{\theta}^{2}(c_{\theta}-3\zeta)+\frac{16c_{\theta}%
^{3}\left(  2c_{\theta}^{2}+c_{\theta}(2-3\zeta)\lambda-3\zeta\lambda\right)
}{2c_{\theta}^{2}+\sqrt{6c_{\theta}^{2}(c_{\theta}+1)\zeta-4c_{\theta}^{3}}%
}-\frac{16(c_{\theta}+1)c_{\theta}^{4}(2c_{\theta}-3\zeta)(c_{\theta}+3\zeta
)}{\left(  2c_{\theta}^{2}+\sqrt{6c_{\theta}^{2}(c_{\theta}+1)\zeta
-4c_{\theta}^{3}}\right)  ^{2}}\right)  }{16c_{\theta}^{2}(3(c_{\theta
}+1)\zeta-2c_{\theta})(c_{\theta}+1)^{-1}},
\end{align*}%
\begin{align*}
e_{1}\left(  \mathbf{x^{1}}\left(  c_{\theta},\zeta,\lambda\right)  \right)
&  =\frac{(c_{\theta}+1)\left(  \lambda^{2}(3(c_{\theta}+1)\zeta-2c_{\theta
})-18\zeta^{2}\right)  }{2(3(c_{\theta}+1)\zeta-2c_{\theta})(c_{\theta}%
\lambda+3\zeta)},\\
e_{2}\left(  \mathbf{x^{1}}\left(  c_{\theta},\zeta,\lambda\right)  \right)
&  =\frac{\lambda(2(c_{\theta}+1)\lambda-c_{\theta}+1)-6\zeta}{2c_{\theta
}\lambda+6\zeta},
\end{align*}
and%
\begin{align*}
e_{1}\left(  \mathbf{x^{2}}\left(  c_{\theta},\zeta,\lambda\right)  \right)
&  =-\frac{3\zeta(c_{\theta}(\lambda-1)+\lambda+1)+2c_{\theta}\lambda
}{2(2\lambda+1)(3(c_{\theta}+1)\zeta-2c_{\theta})}+\frac{\Delta}%
{2(2\lambda+1)(3(c_{\theta}+1)\zeta-2c_{\theta})},\\
e_{2}\left(  \mathbf{x^{2}}\left(  c_{\theta},\zeta,\lambda\right)  \right)
&  =-\frac{\Delta}{2(2\lambda+1)(3(c_{\theta}+1)\zeta-2c_{\theta})}%
-\frac{3\zeta(c_{\theta}(\lambda-1)+\lambda+1)+2c_{\theta}\lambda}%
{2(2\lambda+1)(3(c_{\theta}+1)\zeta-2c_{\theta})},
\end{align*}
where \newline\noindent$\Delta={\surd\left(  3\zeta\left(  -8c_{\theta
}(1+c_{\theta})+15(-1+c_{\theta})^{2}\zeta+72(1+c_{\theta})\zeta^{2}\right)
+\right.  }\newline{2(-1+c_{\theta})\left(  -2c_{\theta}(1+c_{\theta
})+3(1+c_{\theta}(4+c_{\theta}))\zeta-27(1+c_{\theta})\zeta^{2}\right)
\lambda+}\newline{\left(  4c_{\theta}\left(  4+c_{\theta}+4c_{\theta}%
^{2}\right)  -12(-2+c_{\theta})(1+c_{\theta})(-1+2c_{\theta})\zeta
-63(1+c_{\theta})^{2}\zeta^{2}\right)  \lambda^{2}+}\newline{\left.  8\left(
-1+c_{\theta}^{2}\right)  (-2c_{\theta}+3(1+c_{\theta})\zeta)\lambda
^{3}\right)  }.$

\begin{figure}[ptb]
\centering\includegraphics[width=0.7\textwidth]{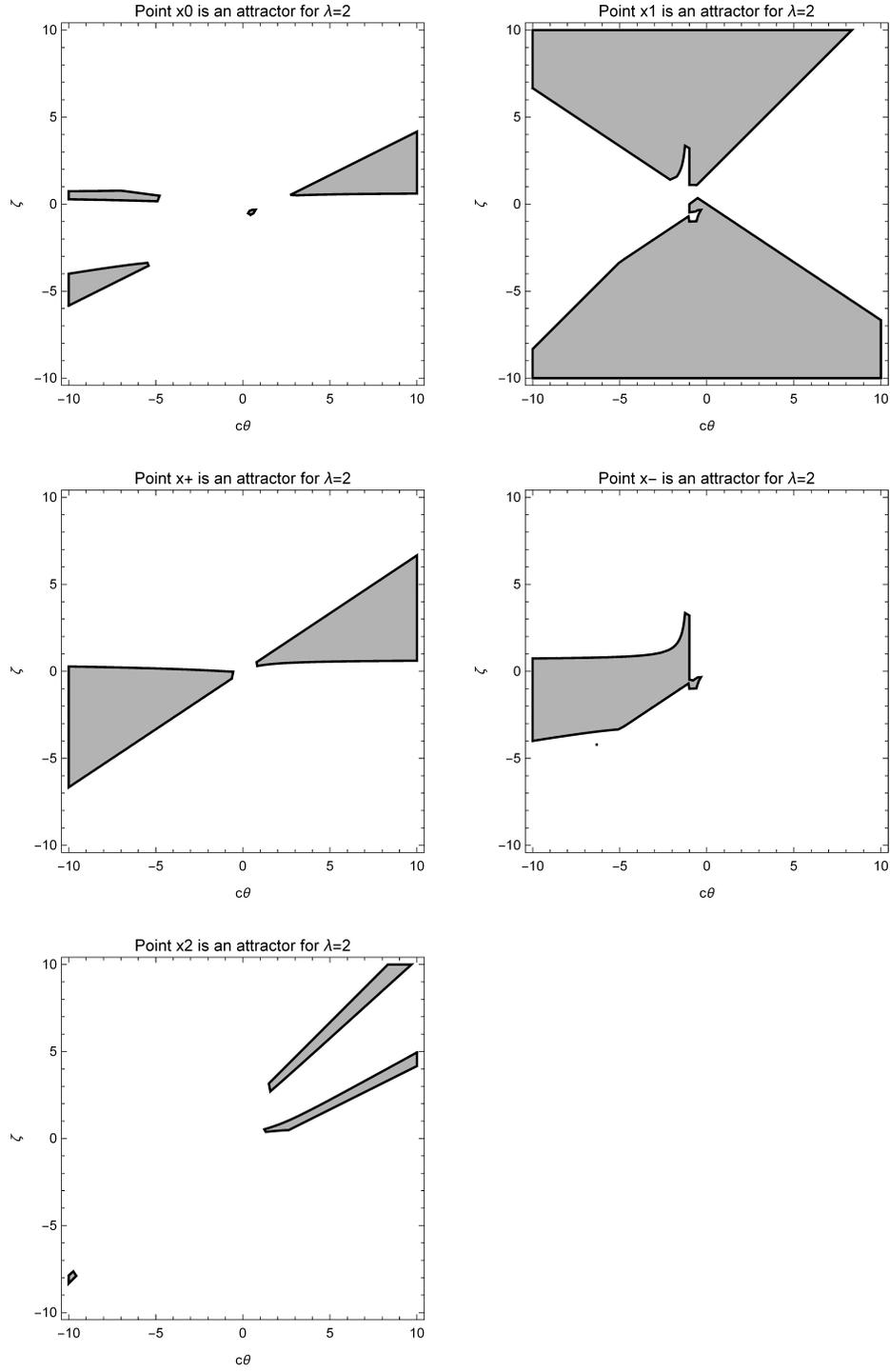} \caption{Region
plots in the space of variables $\left\{  c_{\theta},\zeta\right\}  $, where
the stationary points $\mathbf{x^{0}}$,~$\mathbf{x^{\pm}}$,~$\mathbf{x^{1}}$
and $\mathbf{x^{2}}$ are attractors, for $\lambda=2.$}%
\label{fig6}%
\end{figure}

\begin{figure}[ptb]
\centering\includegraphics[width=0.7\textwidth]{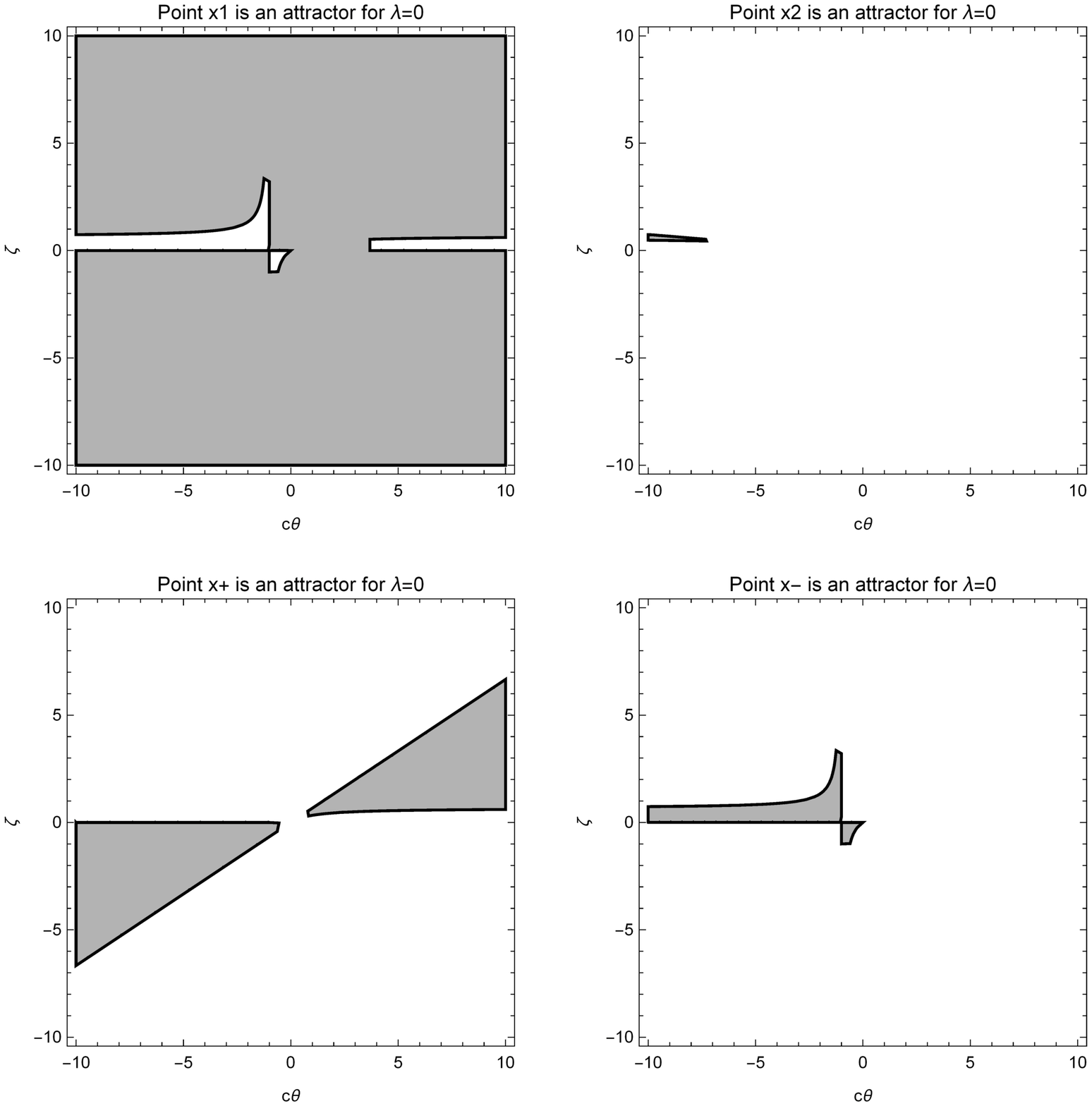}\caption{Region
plots in the space of variables $\left\{  c_{\theta},\zeta\right\}  $, where
the stationary points $\mathbf{x^{\pm}}$,~$\mathbf{x^{1}}$ and $\mathbf{x^{2}%
}$ are attractors, for $\lambda=0.$ Point $\mathbf{x}^{\mathbf{0}}$ always
describes an unstable solution. }%
\label{fig7}%
\end{figure}

In Fig. \ref{fig6} and \ref{fig7} we present the regions in the space of
variables $\left\{  c_{\theta},\zeta\right\}  $ where the stationary points of
the dynamical system (\ref{r01}), (\ref{r02}) are attractors for $\lambda=2$
and $\lambda=0$.

\section{Conclusions}

\label{sec6}

In this work, we considered the extension of Einstein-aether theory in Weyl
geometry. In particular, we considered the Einstein-aether action integral in
the Weyl integrable geometry, where a geometric scalar field is introduced.
The scalar field plays a significant role in the geometry since it defines the
deviation of the Weyl affine connection from that of the Levi-Civita connection.

For the action integral we consider in an isotropic and homogeneous (FLRW)
background spacetime, we observe that the scalar which defines the Weyl affine
connection is introduced into the gravitational field equation and it is
dynamically coupled to the aether field. The scalar field is introduced
two-fold from the Weyl Ricci scalar, and from the symmetric component of the
covariant derivatives for the aether field; from the latter terms, the
coupling then follows. This approach is an alternative way to create an
Einstein-aether scalar field cosmological model. In the case of a
spatially-flat vacuum FLRW spacetime, the field equations admit an exact
solution where the scale factor is power-law and the parameter for the
equation of state can describe acceleration.

We studied the cosmological models where a scalar field potential is included
in the field equations, or dust fluid source contributes to the gravitational
field equations, and when these two terms exist together. For these three
additional systems, we study the dynamics and we find all the possible
asymptotic behaviour of the field equations. We demonstrate our results by
presenting some specific applications. We remark that the results of this work
differ from the previous studies on the Einstein-aether scalar field
cosmological models; however, it is clear that this model can describe an
alternative inflationary behaviour. In a future work, we plan to study the
stability of the small inhomogeneities for this scalar field model.

\begin{acknowledgments}
AP \& GL were funded by Agencia Nacional de Investigaci\'{o}n y Desarrollo -
ANID through the program FONDECYT Iniciaci\'{o}n grant no. 11180126.
Additionally, GL is supported by Vicerrector\'{\i}a de Investigaci\'{o}n y
Desarrollo Tecnol\'{o}gico at Universidad Catolica del Norte. JDB is supported
by the STFC of the United Kingdom.
\end{acknowledgments}

\end{document}